\documentstyle[12pt]{article}
\def\be{\begin{equation}}
\def\ee{\end{equation}}
\def\bea{\begin{eqnarray}}
\def\eea{\end{eqnarray}}

\def\b{\beta}

\def\d{\delta} 
\def\e{\epsilon}
\def\l{\lambda}
\def\pa{\partial}
\def\L{\Lambda}

\author{Hans - J\"urgen Schmidt}

\title{Comparing selfinteracting scalar fields
 and $R + R^3$ cosmological models}

\date{}
\begin{document}
\maketitle

\centerline{Universit\"at Potsdam, Institut f\"ur Mathematik, Am
Neuen Palais 10} 
 \centerline{D-14469~Potsdam, Germany,  E-mail:
 hjschmi@rz.uni-potsdam.de}

\begin{abstract}
We generalize the well-known analogies between $m^2 \phi^2$ 
 and $R + R^2$  theories to include   the selfinteraction 
$ \l \phi^4$-term for the scalar field. It turns out to be the $R + R^3$
 Lagrangian which gives an appropriate model for it. Considering a 
spatially 
flat Friedman cosmological model, common and different properties 
of  these models are discussed, e.g., by linearizing around 
a ground state the masses of the resp. spin 0-parts coincide. 
Finally, we prove   a general conformal equivalence
 theorem between a Lagrangian
$ L = L(R)$, $ L'L'' \ne  0$, and a minimally coupled scalar
 field in a general potential.
\end{abstract}


\section{Introduction}

For the gravitational Lagrangian
\be
     L = (R/2 + \b R^2)/8\pi G \,  ,    
\ee
$R=R_{\rm  crit} = - 1/4\b $
 is the critical value of the curvature scalar 
(cf.  NARIAI 1973, 1974 and SCHMIDT 1986) defined by 
$\pa L/ \pa R = 0$. 
 In regions where  $R/ R_{\rm  crit} < 1$ holds, 
we can define 
 $\psi  = \ln (1 - R/R_{\rm  crit})$  and 
$$
 \tilde g_{ij}   =   (1 - R/R_{\rm  crit})   g_{ij}
$$
 and obtain a Lagrangian
 ($8\pi G = 1$)
\be
 \tilde   L = \tilde R/2
- 3 \tilde g^{ij} \psi_{;i} \psi_{;j}/4 - \left(1
 - e^{-\psi}\right)^2 / 16 \b
\ee
being equivalent to $L$,
 cf. WHITT (1984),  and SCHMIDT (1986) 
for the version of this equivalence used here.

\bigskip

For $\b < 0$, i.e., the absence of 
tachyons in $L$ (1), we have massive 
gravitons of mass $m_0 = (-12 \b)^{-1/2}$
 in $L$, cf. STELLE (1977). For the weak field limit, the potential
 in (2) can be simplified to be $\psi^2/16\cdot \b$,
 i.e., we have got a minimally coupled scalar field whose mass 
is also $m_0$.  (The superfluous factor 
3/2 in (2) can be absorbed by a redefinition of $\psi$.)  
Therefore, it is not astonishing, that all results concerning 
the weak field limit for both $R + R^2$-gravity without 
tachyons and Einstein gravity with a minimally coupled 
massive scalar field exactly coincide. Of course, one cannot 
expect this coincidence to hold for the non-linear region, too, 
but it is interesting to observe which properties hold there also.

\bigskip

We give only one example here: we consider a cosmological 
model of the spatially flat Friedman type, start integrating 
at the quantum boundary (which 
is obtained by  
$$
R_{ijkl}R^{ijkl}
$$
 on the one hand, and $T_{00}$
 on the other hand, to have Planckian values)
 with uniformly distributed initial conditions and look 
whether or not an inflationary phase of 
the expansion appears. In both 
cases we get the following result: The probability $p$
 to have sufficient inflation 
is about $p = 1 - \sqrt{\l} m_0/m_{\rm Pl}$,  i.e., 
$p = 99.992 \% $ if we take $m_0 = 10^{-5} m_{\rm Pl}$ 
 from GUT and $\l = 64$, where $e^\l$
is the linear multiplication factor of 
inflation.\footnote{Cf.  BELINSKY et al. (1985) 
for the scalar field and SCHMIDT (1986) for $R + R^2$, resp.}

\bigskip

From Quantum field theory, however, instead of the massive 
scalar field, a Higgs field with selfinteraction turns out to 
he a better candidate for describing effects of  
the early universe. One of the advances of 
the latter is its possibility to describe a 
spontaneous breakdown of symmetry. In the following, we 
try to look for a purely geometric model for this 
Higgs field which is analogous to the above 
mentioned type where $L = R + R^2$  modelled a massive scalar field.

\section{The Higgs field}

For the massive scalar Field $\phi $ we have
$$
L_m = - \left(
 \phi_{;i} \phi^{;i} - m^2 \phi^2
\right) /2 \qquad  \qquad          (3a)
$$
and for the Higgs field to be discussed now,
$$
     L_\l = - \left(
 \phi_{;i} \phi^{;i} + \mu^2 \phi^2 - \l \phi^4/12 
\right) /2 \qquad  \qquad     (3b)
$$
The ground states are defined by $\phi =$ const.,  
$\pa L/\pa \phi =0$, i.e., $\phi = 0$  for the scalar field, and 
$\phi = \phi_0 = 0$, $\phi = \phi_\pm = \pm \sqrt{6 \mu^2/\l}$
  for the Higgs field.
\be
     \left( + \pa^2 L/ \pa \phi^2 \right)^{1/2}
\ee
is die effective mass at these points, i.e., $m$ for the 
scalar field (3a), and $i\mu$  at $\phi = 0$  and 
 $\sqrt{2} \mu $
 at   $\phi = \phi_\pm$
 for the Higgs field (3b). To have a vanishing 
Lagrangian at the ground state     $ \phi_\pm$
we add  a   constant
\be
\L  = -3 \mu^4/2 \l 
\ee
to the Lagrangian
 (3b). The final Lagrangian reads
\be
L = R/2 + L_\l + \L 
\ee
 with $L_\l$ (3b) and $\L$ (4).

\section{The nonlinear gravitational Lagrangian}

Preliminarily  we direct the attention to the 
following fact: on the one hand, for Lagrangians 
(3a,b) and (5)
the transformation $\phi \to - \phi$
 is a pure 
gauge transformation, it does not change any 
invariant or geometric objects. On the other hand,
\be
 R_{ijkl} \to -  R_{ijkl}
\ee
or simpler
\be
R \to  -R
\ee
is a gauge transformation at 
the linearized level only: taking
$$
g_{ik} = \eta_{ik} +  \e h_{ik} \,   ,
$$
where
$$
\eta_{ik} = {\rm diag} (1, -1, -  1, - 1)
$$
then $\e \to - \e $ implies (6) at the linearized level in $\e$ 
whereas even (7) does not hold quadratically in $\e$. 
This corresponds to  fact that the $\e^2$-term in (2) 
(corresponding to the $\psi^3$-term in the development
 of $\tilde L$ in powers of $\psi$) 
  is the first one to break  the $\psi \to - \psi$ symmetry in (2).

\bigskip

Now, let us introduce the general nonlinear 
Lagrangian $L = L(R)$  which we at the moment 
only assume to be an analytical function of $R$.
 The 
ground states are defined by $R =$ const., i.e.,
\be
     L'R_{ik}- g_{ik} L/2   = 0    \, .
\ee
Here, $L' = \pa L/ \pa R$.

\subsection{Calculation of the ground states}

From eq. (8) one immediately sees that 
$\pa L/ \pa R =0 $ 
defines critical values of the curvature scalar. 
For these values $ R = R_{\rm  crit}$    it holds: 
For $ L(R_{\rm  crit})  \ne  0 $ 
 no such ground state exists,  and for 
$ L(R_{\rm  crit}) =  0 $, we have  only one 
equation 
 $ R = R_{\rm  crit}$ 
 to be solved with  10 
arbitrary functions $g_{ik}$. 
 We call these ground states 
degenerated ones.  For $L = R^2$, 
 $ R_{\rm  crit}
 = 0$, this has 
been discussed by BUCHDAHL  (1962). 
 Now, let us concentrate on the case $\pa L/ \pa R  \ne  0$. 
 Then $R_{ij} $  is proportional to $g_{ij}$  with a constant 
proportionality factor, i.e., each ground 
state is an Einstein space
\be
R_{ij}= R g_{ij} /4 \,    , 
\ee    
 with a prescribed constant value $R$. 
Inserting (9) into (8) we get as condition for ground states
$$
            RL' = 2L \, .
$$
     As an example, let $L$  be a third order polynomial
\be
            L = \L  + R/2 + \b R^2 + \l R^3/12 \,  . 
\ee
We consider only Lagrangians with a positive 
linear term as we wish to reestablish Einstein gravity in 
the $\L \to   0$  weak 
field limit, and $\b  <  0$  to exclude tachyons there.

\bigskip

For $\l = 0$ we have (independently of $\b$!) the only 
ground state $R = -4\L$. It is a degenerated one 
if and only if   $\b \L = 1/16$. That implies that for 
usual $R + R^2$ gravity (1) \  ($\l = \L = 0$) \ $ R = 0 $  is the 
only ground state and it is a nondegenerated one.

\bigskip

Now, let $\l \ne 0$ and $\L =  0$. To get nontrivial 
ground states we further 
require $\l  > 0$. Then, besides $R = 0$, the ground states are
\be
     R=R_\pm  = \pm \sqrt{6/\l}
\ee
being quite analogous to those of the Higgs 
field (3b). The ground state $R = 0$  is  not 
degenerated (of course, this statement  is 
independent of $\l$  and holds true, as one knows, 
for $\l = 0$.) To exclude tachyons, we require 
$\b < 0$, then $R_-$  is not degenerated and 
$R_+$  is degenerated if and only if   $\b  = -\sqrt{6/\l}$.

\bigskip

The case $\l \L  \ne  0$  will not be considered here.

\subsection{Definition of the masses}

For the usual $R + R^2$  theory (1), the mass is 
$$
m_0 = (R_{\rm  crit}/3)^{1/2} =  (- 12 \b)^{1/2} \, .
$$
But how to define the graviton's masses for the
 Lagrangian (10)? To give such a definition a 
profound meaning one should  do the following: 
linearize the full vacuum field equation around 
the ground state (preferably de Sitter- or anti-de Sitter space, 
 resp.) decompose its solutions with respect to 
a suitably chosen orthonormal system (a 
kind of higher spherical harmonics) and look for the 
properties of its single modes. For $L$  (1) this procedure just gave 
 $m_0$.

\bigskip

A little less complicated way to look at this mass problem 
is to consider a spatially flat Friedman cosmological model 
and to calculate the frequency with which the scale factor 
oscillates around the ground state, from which the 
mass $m_0$  turned out to be the graviton's mass for $L$  (1), too.

\bigskip

Keep in mind, 1. that all things concerning a 
linearization around flat vacuous space-time do 
not depend on the parameter $\l$ neither for the 
Higgs field nor for the $L(R)$  model, and 2. 
that a field redefinition $R \to   R^\ast + R_\pm$  is {\it  not} 
possible like 
 $\phi \to \phi^\ast + \phi_\pm$ 
 because curvature remains    absolutely present.

\section{The cosmological model}

Now we take as Lagrangian eq. (10) and as line element
\be
 ds^2= dt^2 - a^2(t) (dx^2 + dy^2 + dz^2)   \, .
\ee
The dot denotes $d/dt$  and $h = \dot a/a$. We have
\be
          R =  -6 \dot h - 12h^2 \,   ,
\ee
and the field equation will be obtained as follows.

\subsection{The field equation}

For $L = L(R)$  the variation
$$
\d \left( L \sqrt{-g} \right) / \d g^{ij} =0
$$
 gives with $L' = \pa L/ \pa R $
\be
L' R_{ij} - g_{ij} L/2 + g_{ij } \Box L' - L'_{;ij}
=0                  \, ,
\ee
cf. e.g., NOVOTNY (1985). (Be aware of sign 
 errors in the paper
 of KERNER (1982) such  that the results of it are wrong.
Nevertheless, his ideas are fruitful ones.) It holds 
\be
L'_{;ij} = L'' R_{;ij} + L''R_{;i} R_{;j} \, .
\ee
With eq. (15), the trace of eq. (14) reads
\be
     L'R - 2L + 3L'' \Box  R + 3L''
R_{;k}R^{;k} = 0 \,   ,  
\ee

i.e., with $L$  eq. (10)
$$
-2 \L 
-R/2 + \l R^3/12 + 6 \b \Box R
 + \frac{3 \l}{2}
(R \Box R +
R_{;k}R^{;k} )
=0 \,  .  
$$
Inserting eqs. (12), (13), (15) 
into the $00$-component of eq. (14) we get the equation
\be
0    = h^2/2 - \L /6 
- 6 \b
(2h \ddot h - \dot  h^2 + 
6h^2 \dot h )  + 3 \l (\dot h  + 2h^2)
 (6h \ddot h  
+ 19h^2 \dot h - 2\dot h^2 - 2h^4) \,  . 
\ee
The remaining components are a consequence of this one.

\subsection{The masses}
Linearizing the trace equation (16) around the flat
 space-time (hence, $\L = 0$) gives (independently of 
$\l$, of course) $ R = 12 \b \Box R$, and the oscillations around 
the flat space-time indeed correspond to a mass $m_0
  = (-12 \b) ^{ -1/2}$. 

\bigskip

Now, let us linearize the ground 
states (11) by inserting $\L = 0$
 and $R =\pm\sqrt{6/\l} + Z$  into eq. (16). It gives
$$
Z = \left( -6\b \mp \sqrt{ 27 \l /2} \right) \Box Z \,  ,
$$
and, correspondingly,
\be
     m_\pm = \left(  6 \b \pm \sqrt{ 27 \l /2} \right)^{-1/2}   \, .
\ee
For $\b \ll  - \sqrt \l$, \   $m_\pm$ 
is imaginary, and its 
absolute value differs by a factor $\sqrt 2$ from $m_0$. 
This is quite analogous to the 
$\l \phi^4$-theory, cf.  
sect. 2. Therefore, we concentrate on discussing 
this range of  parameters.

\bigskip

For the ground state for $\L  \ne  0$, $\l = 0$ we get with 
$
R = -4 \L  + Z $  just $ Z = 12\b \Box  Z$, i.e., mass $m_0$ 
 just as in the case $\l  = \L = 0$.

\bigskip

Let us generalize this estimate to 
 $L = L(R)$; $R = R_0 = $ const. is a ground 
state if    
$$
 L'(R_0) R_0 = 2L(R_0)
$$
  holds. It is degenerated if 
$L'(R_0) = 0$. Now, linearize around 
$R = R_0$: $R = R_0 + Z$. For $L''(R_0) = 0$, 
only $Z = 0$ solves the linearized equation, and 
 $R = R_0$  is a singular solution. For $L''(R_0)  \ne  0$ we 
get the mass
\be
m = \left( R_0/3 - L'(R_0)/3L' (R_0 ) \right)^{1/2} 
\ee
meaning the absence of tachyons 
for real values $m$. Eq. (19) is the analogue to eq. (3).

\subsection{The Friedman  model}

Here we only consider the spatially flat 
Friedman model (12). Therefore, we can 
discuss only de Sitter  stages 
with $R < 0$ , esp. the ground state 
 $R_+$ eq. (11) does not enter our discussion but $R_-$ does. 

\bigskip

Now, let $\L =  0$. Solutions of eq. (17) with 
constant values $h$ are $h=  0$ 
(flat space-time) and 
$$
h = \frac{1}{
\sqrt[4]{24 \l} }
$$
(de Sitter space-time) representing the 
non-degenerated ground states $R = 0$ 
and $ R = R_-   = - \sqrt 6/\l$,
 resp. Eq. (17) can be written as
\bea
     0 = h^2(1 - 24 \l h^4)/2 + h \ddot h \left(
 1/m_0^2 + 18 \l (\dot h + 2 h^2) \right) \nonumber \\
 - 6 \l \dot h^3 + \dot h^2 (45 \l h^2 - 1/2m_0^2 )
+ 3 h^2 \dot h (1/m_0^2 + 36 \l h^2 ) \, .
\eea
First, let us consider the singular curve 
defined by the vanishing of the 
coefficient of   $\ddot h$  in eq. (20) in 
the $h - \dot h$-phase plane. It is, besides $h = 0$, the curve
\be
\dot h =    -2h^2 - 1/18\l m_0^2
\ee
 i.e., just the curve 
$$
R = 1/3\l m_0^2 = -4 \b /\l
$$
 which is 
defined by $L'' = 0$, cf.  eq. (16). This value
 equals $R_+ $ if $\b  = - \sqrt{3 \l /8}$, 
this value we do not discuss here. Points 
of the curve (21) fulfil eq. (20) for
$$
h = \pm 1/18\l m_0^3 \sqrt 3 \sqrt{1-1/18\l m_0^4} 
$$
only, which is not real because of $\l \ll m_0^4$.

\bigskip

Therefore, the space of solutions is 
composed of at least 2 connected 
components. 

\bigskip

Second, for $h = 0$ we have $\dot h = 0$ or
\be
\dot h = - 1/12 \l m^2   \, .
\ee
$h = \dot h =0$
 implies $h \ddot h \ge 0$,  i.e. $h$ does not 
change its sign. (We know this already 
from M\"ULLER and SCHMIDT (1985), 
where the same model with 
 $\l = 0$ is discussed.) In a neighbourhood of (22) we can 
make the ansatz
$$
h = -t/12 \l m_0^2 + \sum_{n=2}^\infty a_n t^n
$$
which has solutions with 
arbitrary values $a_2$. That means:  one can 
change from expansion to subsequent recontraction,  
but only through 
the ``eye of a needle" (22). On the other hand, 
a local minimum of the scale factor 
never appears. Further, (22) does not 
belong to the connected component of  flat space-time.

\bigskip

But we are especially interested in the latter 
one, and therefore, we restrict to the subset 
$\dot h > \dot h$(eq. (21)) and need 
only to discuss expanding 
solutions $h \ge 0$. Inserting $\dot h = 0$,
$$
\ddot h =    h(24 \l h^4 - l)/(2/m_0^2 + 72\l h^2)
$$
turns out, i.e., $\ddot h > 0$ for $h > 1
/ \sqrt[4]{24 \l }$
 only. 
All other points in the $h - \dot h$ phase 
plane are regular ones, and one can 
write $d \dot h / dh \equiv \ddot h / \dot h  = F(h, \dot h)$
 which can be calculated by eq. (20).

\bigskip

For a concrete discussion let $ \l \approx 10^2 l^4_{\rm Pl} $ 
 and $m_0 = 10^{-5}m _{\rm Pl} $.  Then both 
conditions 
$\b \ll - \sqrt \l$
 and $\vert  R_- \vert < l^{-2}_{\rm Pl}$
 are fulfilled.
 Now the qualitative behaviour of 
the solutions can be summarized: There 
exist two special solutions which approximate the 
ground state $R_-$ for $t \to - \infty$. All other 
solutions have a past singularity $h \to \infty$.  
Two other special solutions approximate 
the ground state $R_-$ for $t \to   + \infty$. 
Further solutions have a future 
singularity $h \to \infty$, and all other solutions 
have a power like behaviour for $t \to \infty$,
 $ a(t)$  oscillates around the 
classical dust model $a(t) \sim t^{2/3}$. 
But if we restrict the initial conditions to lie in 
a small neighbourhood of the unstable ground 
state $R_-$, only one of the following three cases appears:

\noindent 
1.   Immediately one 
goes with increasing values $h$ to a singularity.

\noindent 
2.   (As a special case)
 one goes back to the de Sitter stage $R_-$.

\noindent
3.   (The only interesting one) 
One starts with a finite 
$l_{\rm Pl}$-valued inflationary era, goes 
over to 
a GUT-valued second inflation and ends with a power-like 
Friedman  behaviour.

\bigskip

In the last case to 
be considered here, let $\l = 0$, $\L > 0$ and $\b < 0$. 
The analogue to eq. (20) then reads
$$
0    = h^2/2 - \L /6 + (2h \ddot h - \dot  h^2 +
 6h^2 \dot h)/2m_0^2 \,  .
$$
Here, always $h \ne 0$  holds, we consider 
only expanding solutions $h > 0$. For $\dot h = 0$ we have
$$
\ddot h 
=    (\L  m_0^2/3 -m_0^2 h^2)/2h \, .
$$
For $\ddot h = 0$ we have $\dot h >  m_0^2/6$ and
$$
h = (\L /3 + \dot  h^2/m_0^2)^{1/2}
 (1 + 6 \dot h/m_0^2)^{-1/2} \, .
$$
Using the methods of 
M\"ULLER and SCHMIDT (1985) (where the case $\L = 0$
 has been discussed) we obtain the following result:
All solutions approach the de Sitter phase $h^2 = \L /3$
 as $t \to \infty$.
 There exists one special solution  approaching 
$ \dot h = -m_0^2/6$ for
$ h \to \infty $,
 and all solutions have a past singularity $h \to \infty$.
 For a sufficiently small value $\L$
 we have again two different inflations in most of all models.

\section{The generalized equivalence}

In this section we derive a general 
equivalence theorem between a nonlinear 
Lagrangian $ L(R)$  and a minimally coupled 
scalar field $\phi$ with a general 
potential with Einstein's theory. Instead of $\phi$  we take

$$
\psi = \sqrt{2/3} \, \phi\, .
$$
This is done to avoid square roots in the exponents. 
Then the Lagrangian for the scalar field reads
\be
 \tilde   L = \tilde R/2
- 3 \tilde g^{ij} \psi_{;i} \psi_{;j}/4 
 + V(\psi) \, .
\ee
At ground states $\psi = \psi_0$, defined by
 $\pa V/ \pa \psi = 0$  the effective mass is
\be
          m= \sqrt{2/3}
 \sqrt{ \pa^2 V/ \pa \psi^2   }
   \,  ,   
\ee
cf. eq. (3). The variation $ 0 = \d \tilde L/ \d \psi$ gives
\be
         0 =\pa V/ \pa \psi + 3 \  \tilde{} \  \Box \psi / 2
\ee
and Einstein's equation is
\be
\tilde E_{ij} = \kappa \tilde T_{ij}
\ee
with
\be
\kappa \tilde T_{ij} =  3 \psi_{;i}\psi_{;j} /2 +
   \tilde g_{ij} \left(
 V(\psi) - \frac{3}{4} \tilde g^{ab} \psi_{;a}\psi_{;b}
\right)        \, .
\ee
Now, let
\be
   \tilde g_{ij} =e^\psi  g_{ij} \, .
\ee
The conformal transformation (28) shall 
be inserted into eqs. (25, 26, 27). One obtains from (25) with 
\bea
\psi^{;k} := g^{ik} \psi_{;i}
     \nonumber \\
\Box \psi + \psi^{;k}\psi_{;k}
= - 2 (e^\psi \pa V/\pa \psi )/3
\eea
and 
from
 (26, 27)
\be
E_{ij} = \psi_{; ij} + \psi_{;i}\psi_{;j} +g_{ij} 
\left(
e^\psi V(\psi) - \Box \psi
 - \psi_{;a} \psi^{;a}
\right) \, .
\ee
Its trace reads
\be
        -R = 4 e^\psi  V(\psi) - 3 \Box \psi 
 - 3 \psi_{;a} \psi^{;a} \, .
\ee
Comparing with eq. (29) one obtains
\be
        R = R(\psi) =
  - 2e^{-\psi} \pa \left( e^{2\psi} V(\psi) \right) /
 \pa \psi \, .
\ee
Now, let us presume $\pa R/\pa \psi  \ne  0$,
 then eq. (32) can be inverted as
\be
 \psi = F(R) \,   . 
\ee 
In the last step, 
eq. (33) shall be inserted into eqs. (29, 30, 31). Because of
$$
F(R)_{; ij}
= \pa F/ \pa R \cdot  R_{; ij} + \pa^2 F/ \pa R^2 \cdot R_{; i}R_{;j}
$$
and $\pa F/\pa R  \ne  0$,  eq. (30) is a fourth 
order equation for the metric $g_{ ij}$.   We 
try  to find a Lagrangian    $L = L(R)$ such 
that the equation $ \d L \sqrt{-g} / \d g^{ij}
=  0$ 
becomes just eq. (30). For $L' = \pa L/\pa R  \ne  0$,
 eq. (14) can be solved to be
\be
E_{ij} = - g_{ij}R/2 + g_{ij}L/2 L'  - g_{ij}
\Box L'/L' 
  - L'_{;ij}/L' \, . 
\ee
We compare     the coefficients of the $R_{;ij}$  terms  
in eqs. (30) and (34), this gives 
\bea
\pa F/\pa R = L''/L' \,  , \qquad {\rm  hence} \nonumber   \\
L(R) = \mu \int_{R_0}^R e^{F(x)} dx + \L_0\, ,
\eea 
with suitable constants $\L_0$, $\mu$, and $R_0$, $\mu  \ne  0$.
 We fix them as follows: We are interested in a neighbourhood of 
$R = R_0$
  and require $L'(R_0) = 1/2$.
 (Otherwise $L$
 should be multiplied 
by a constant factor.) Further, a constant translation
of $\psi$  can be used to obtain $F(R_0) = 0$,
 hence $\mu = 1/2$, $L(R_0) = \L_0$, and
$$
L' (R_0) 
= \pa F/\pa R(R_0)/2  \ne  0\,  .
$$
With (35) being fulfilled, the traceless 
parts of eqs. (30) and (35) 
identically coincide. Furthermore, we have
$$
\Box L'/L' = \Box F + F^{;i} F_{;i}
$$
and it suffices to test the validity of the relation
$$
e^F V(F(R)) = -R/2 + L/2L' \, . 
$$
It holds
\bea
2L' = e^F\,  , \qquad {\rm i.e.,} \nonumber \\
 e^{2F} V(F(R)) = L - R e^F/2 \, .
\eea
At $R = R_0$, this relation reads $V(0) = \L_0 - R_0/2$.  
Applying $\pa /\pa R$  to eq. (36) gives 
just eq. (29), and, by  the way, $V'(0) =
R_0/2 -    2\L_0$. In sum,
$$
L(R) = V(0) + R_0/2 + 
 \int_{R_0}^R e^{F(x)} dx/2 \, ,
$$
where $F(x)$ is defined via $F(R_0) = 0$,
$$
\psi
 = F\left( -2 e^{-\psi}
\pa(e^{2\psi} V(\psi))
 / \pa \psi 
 \right) \, .
$$

\bigskip

Now, let us go the other direction: 
Let $L = L(R)$ be given 
such that  at $R = R_0$,
 $ L'L''  \ne  0$. By a constant change of   $L$ let 
$L'(R_0) =
    1/2$. 
Define $\L_0 = L(R_0)$, $\psi = F(R) = \ln (2L'(R))$  and 
consider the inverted function $R = F^{-1}(\psi)$.  Then
\be
V(\psi) = (\L_0 - R_0/2) e^{-2\psi}
- e^{-2\psi} \int_0^\psi
 e^x \ F^{-1}(x) dx/2
\ee
is the potential ensuring the 
above mentioned conformal equivalence. 
This procedure is possible at all $R$-intervals 
where $L'L''  \ne  0$ holds. For analytical 
functions $L(R)$, this inequality can be violated 
for discrete values $R$  only (or one has simply 
a linear function $L(R)$ being Einstein gravity with $\L$-term).

\bigskip

\noindent
{\it Examples:} \  1. Let $L = \L + R^2$, $R_0 = 1/4$,
  then $4R = e^\psi$  and
\be
V(\psi ) = \L  e^{-2\psi} - 1/16\, .
\ee
(For $\L = 0$, this is proven in 
BICKNELL (1974) 
and STAROBINSKY and SCHMIDT (1987).)

\bigskip

\noindent
2.   Let $L = \L + R/2 + \b R^2 + \l 
R^3/12$, \  $R_0=0$, hence $\b  \ne  0$ is necessary. We get
\bea
e^\psi -    1 = 4 \b R + \l R^2/2 \qquad {\rm   and} \nonumber \\
V(\psi) = \L e^{-2\psi} +  \nonumber \\ 
 2\b \l^{-1} e^{-2\psi}
 \left( e^\psi - 1 - 
16\b^2(3\l)^{-1}
 ((1 + \l(e^\psi - l)/8\b^2)^{3/2} - 1) \right) \, . 
\eea
The limit $\l \to  0$  in eq. (39) is  possible and leads to
$$
V(\psi) = \L e^{-2\psi} -
 (e^{-\psi}- 1)^2/16\b \, , \qquad {\rm  cf. \  eq. \,  (2).}
$$

\bigskip
  
Now, let $R_0$ be a non-degenerated ground state, hence
$$
L(R) = \L_0 + (R - R_0)/2 + L''(R_0) (R - R_0)^2/2 +\dots 
$$
with $L''(R_0)     \ne  0$  and $\L_0 = R_0/4$, 
cf. sct. 3.1. Using eq. (37) we get $V'(0) = 0$  and 
$$
V''(0) = R_0/2 - 1/4L''(R_0) \, .
$$
Inserting this into eq. (24) 
we exactly reproduce eq. (19). 
This fact once again confirms 
the estimate (19) and, moreover, shows 
it to be a true analogue to eq. (3). 
To understand this coincidence 
one should note that at ground states, 
the conformal factor becomes a constant = 1. \par
\noindent 
The author gratefully acknowledges 
an initiating idea of  Dr. U. KASPER.

\section*{References}

\noindent 
BICKNELL, G.: 1974, J. Phys. A {\bf  7}, 1061.

\noindent 
BELINSKY, V. A., GRISHCHUK, L. P.,
 KHALATNIKOV, I. M., ZELDOVICH, Ya. B.: 1985, Phys. Lett. B 
{\bf 155}, 232.

\noindent 
BUCHDAHL, H.: 1962, Nuovo Cim. {\bf 23}, 141.

\noindent 
KERNER, R.: 1982, Gen. Rel. Grav. {\bf  14}, 453.

\noindent 
M\"ULLER, V., SCHMIDT, H.-J.: 1985, Gen. Rel. Grav. {\bf  17}, 769.

\noindent 
NARIAI, H.: 1973, Progr. Theor. Phys. {\bf  49}, 165.

\noindent 
NARIAI, H.: 1974, Progr. Theor. Phys. {\bf 51}, 613.

\noindent 
NOVOTNY, J.: 1985, Coll. J. Bolyai Math. Soc. Budapest.

\noindent 
SCHMIDT, H.-J.: 1986, Proc. Conf.  GR 11 Stockholm, 
 p. 117, and Thesis B, Academy of Sciences Berlin, GDR.

\noindent 
STAROBINSKY, A. A., SCHMIDT, H.-J.: 1987, 
Class. Quant. Grav. {\bf  4}, 695.

\noindent 
STELLE K.: 1977, Phys. Rev. D {\bf  16}, 953.

\noindent 
WHITT, B.: 1984, Phys Lett. B {\bf  145}, 176.

\bigskip

Received 1986 November 5

\medskip

\noindent 
{\small {This is a  reprint from Astronomische Nachrichten, 
done with the kind permission of the copyright owner, only some obvious
 misprints have been cancelled;    
 Astron. Nachr. {\bf 308} (1987) Nr. 3, pages 183 - 188;  
  Author's address that time:  
Zentralinstitut f\"ur  Astrophysik der AdW der DDR, 
1591 Potsdam, R.-Luxemburg-Str. 17a.}}

\end{document}